\documentclass[acmsmall]{acmart}

\AtBeginDocument{%
  \providecommand\BibTeX{{%
    \normalfont B\kern-0.5em{\scshape i\kern-0.25em b}\kern-0.8em\TeX}}}
\usepackage{multirow}

\usepackage{hyperref}  
\usepackage[hyphenbreaks]{breakurl}
\usepackage{caption}
\usepackage{subcaption}
\usepackage{graphicx}
\usepackage{makecell}
\usepackage{booktabs}
\usepackage{array}
\usepackage{mdframed}
\newcolumntype{L}[1]{>{\raggedright\let\newline\\\arraybackslash\hspace{0pt}}m{#1}}
\newcolumntype{C}[1]{>{\centering\let\newline\\\arraybackslash\hspace{0pt}}m{#1}}
\newcolumntype{R}[1]{>{\raggedleft\let\newline\\\arraybackslash\hspace{0pt}}m{#1}}
\setcopyright{rightsretained}
\acmDOI{10.1145/3660784}
\acmYear{2024}
\copyrightyear{2024}
\acmSubmissionID{fse24main-p226-p}
\acmJournal{PACMSE}
\acmVolume{1}
\acmNumber{FSE}
\acmArticle{77}
\acmMonth{7}
\received{2023-09-28}
\received[accepted]{2024-04-16}

\begin{document}
\title{How to Gain Commit Rights in Modern Top Open Source Communities?}

\author{Xin Tan}
\orcid{0000-0003-1099-3336}
\affiliation{%
  \institution{School of Computer Science and Engineering, Beihang University}
  %\streetaddress{1 Th{\o}rv{\"a}ld Circle}
  \city{Beijing}
  \country{China}}
\email{xintan@buaa.edu.cn}

\author{Yan Gong}

\authornote{Both authors contributed equally to this research.}
\email{gongyan020428@163.com}
\orcid{0009-0008-6447-9010}
\author{Geyu Huang}
\authornotemark[1]
\orcid{0009-0008-6260-4201}
\email{20373244@buaa.edu.cn}
\affiliation{%
  \institution{School of Computer Science and Engineering, Beihang University}
  %\streetaddress{P.O. Box 1212}
  \city{Beijing}
  \country{China}
}

\author{Haohua Wu}
\orcid{0009-0006-3976-3816}
\affiliation{%
  \institution{ShenYuan Honors College, Beihang University}
  %\streetaddress{1 Th{\o}rv{\"a}ld Circle}
  \city{Beijing}
  \country{China}}
\email{oliverhaohuawu@buaa.edu.cn}

\author{Li Zhang}
\orcid{0000-0002-2258-5893}
\authornote{Corresponding author.}
\affiliation{%
  \institution{School of Computer Science and Engineering, Beihang University}
  %\streetaddress{1 Th{\o}rv{\"a}ld Circle}
  \city{Beijing}
  \country{China}}
\email{lily@buaa.edu.cn}

%\renewcommand{\shortauthors}{Xin Tan and et al.}




\begin{abstract}
The success of open source software (OSS) projects relies on voluntary contributions from various community roles. Among these roles, being a committer signifies gaining trust and higher privileges in OSS projects. Substantial studies have focused on the requirements of becoming a committer in OSS projects, but most of them are based on interviews or several hypotheses, lacking a comprehensive understanding of committers' qualifications. To address this knowledge gap, we explore both the policies and practical implementations of committer qualifications in modern top OSS communities. Through a thematic analysis of these policies, we construct a taxonomy of committer qualifications, consisting of 26 codes categorized into nine themes, including ``\textit{Personnel-related to Project}'', ``\textit{Communication}'', and ``\textit{Long-term Participation}''. We also highlight the variations in committer qualifications emphasized in different OSS community governance models. For example, projects following the ``\textit{core maintainer model}'' place great importance on project comprehension, while projects following the ``\textit{company-backed model}'' place significant emphasis on user issue resolution. Based on the above findings, we propose eight sets of metrics and perform survival analysis on two representative OSS projects to understand how these qualifications are implemented in practice. We find that the probability of gaining commit rights decreases as participation time passes. The selection criteria in practice are generally consistent with the community policies. Developers who submit high-quality code, actively engage in code review, and make extensive contributions to related projects are more likely to be granted commit rights. However, there are some qualifications that do not align precisely, and some are not adequately evaluated. This study enhances trust understanding in top OSS communities, aids in optimal commit rights allocation, and empowers developers' self-actualization via OSS engagement.
\end{abstract}

%\keywords{open source software, committers, trust, committer immigration}

\maketitle

\section{Introduction}
%Since the 1980s, Open Source Software (OSS) innovations have fundamentally changed software development and realization~\cite{feller2002understanding}. Numerous excellent OSS projects have been produced based on this development mode, including world-class software such as the Linux kernel and Android. Unlike traditional software development, OSS development is driven by collaboration among geographically distributed developers who share common goals and interests~\cite{raymond1999cathedral}. Therefore, the success of OSS projects is inseparable from the communities behind them.

Since the 1980s, Open Source Software (OSS) has revolutionized software development~\cite{feller2002understanding}. Collaborative efforts among geographically dispersed developers have created world-class OSS projects such as the Linux kernel and Android. The success of OSS projects is inseparable from the communities behind them~\cite{raymond1999cathedral}.

Because the source code is transparent, developers with different backgrounds all have opportunities to contribute. While this open development mode can bring continuous innovation~\cite{cole2001continuous}, it may also bring the risk of low quality if without restrictions. However, in fact, OSS is often considered more reliable and secure compared with closed-source software~\cite{bonaccorsi2003open}, mainly because of its unique community structure and the shared trust among community members. To ensure code quality, OSS communities have a hierarchical or onion-like structure~\cite{crowston2003social,raymond1999cathedral}. At the center is a small group of developers (usually called ``committers'') who have permission to commit source code to the main repository and oversee the design and evolution of the project~\cite{sinha2011entering}.\footnote{Developers with commit rights may be referred to as collaborators or project members in some communities. However, we use the term ``committers'' throughout the paper for brevity, although these roles differ slightly.} Although this core group is highly trusted, its members are not static. The dynamic changes of this core group are significant because injecting new blood can facilitate OSS communities to continue to innovate~\cite{goldman2005innovation,steinmacher2016overcoming}. Moreover, external developers are eager to enter this circle of trust (especially the top OSS communities) because it is a great way to improve their technical skills and proof of their social status~\cite{von2012carrots,qiu2019going}. 

However, it is not easy to distribute the commit right in OSS communities~\cite{tan2020scaling,ihara2014early}. To ensure proper distribution, OSS communities need to consider whether a developer is qualified to grant the commit right, including both technical and social aspects~\cite{bird2007open}. Potential committers need to earn respect by demonstrating their personal eminence and earning their peers' trust and confidence. For developers, successful onboarding in top communities is already a challenge~\cite{steinmacher2014hard,tan2020first}, let alone entering their trust circle. Thus, a critical question is ``\textbf{how to select/become committers in OSS communities?}''. This question directly affects whether OSS communities can be sustainable and whether individual developers can realize their value through OSS participation.

Some studies focus on understanding the important factors for being committers~\cite{sinha2011entering, bird2007open, ihara2014early, jergensen2011onion}. However, these studies are just based on interviews or several hypotheses without a comprehensive understanding of committer qualifications, i.e., the requirements that developers need to meet in order to obtain the commit right. Top OSS communities typically have policies about how to apply for the commit right. For example, the Node.js community stipulates that \textit{existing collaborators (i.e., developers with the commit right) can nominate someone to become a collaborator. Nominees should have significant and valuable contributions across the Node.js organization}. Such policies are valuable resources for understanding how trust is established in modern top OSS communities. To this end, building upon the above research, we investigate committer qualifications in top OSS communities. We propose the following research questions to guide our study.
\begin{itemize}
    \item[\textbf{RQ1:}] What qualifications are required for obtaining commit rights in modern top OSS communities? 
    \item[\textbf{RQ2:}] Do the actual selection criteria for committers align with these qualifications in practice?
\end{itemize}

To answer these questions, we investigate 43 popular (with the most stars) GitHub projects that have policies about how to become committers. We find that only around 2\% of external developers obtain the commit right after a period of contribution despite the proportion of committers in communities reaching about 30\%. It indicates that it is a great challenge to gain the trust of top OSS communities. By adopting a thematic analysis of community policies, we identify 26 codes belonging to nine themes (e.g., \textit{Personnel-related to Project}, \textit{Nomination}, and \textit{Long-term Participation}). Combined with four typical OSS governance models, we find that these models have different emphases on committer qualifications. For example, the company-backed model focuses more on user issue resolution, while the core maintainer model pays more attention to project comprehension. With these expected qualifications, we investigate how projects select committers in practice. We designed eight sets of metrics and conducted survival analysis on the two representative projects. We find that the selection criteria in practice are basically consistent with policies. For example, if developers submit high-quality code, actively participate in code reviews, and contribute to projects with a close relationship, they are more likely to be granted the commit right. Some of the differences between projects are related to their community governance models. However, we also noticed that the expected and actual committer qualifications are not exactly consistent and some dimensions are not adequately evaluated in practice.

This paper fills the knowledge gap regarding committer qualifications in top OSS communities through a combination of policy investigation and practical analysis. It enhances our understanding of trust-building mechanisms in modern top OSS communities, facilitates fairer distribution of the commit right, and helps individual developers achieve self-actualization in OSS communities. Overall, this paper makes the following contributions:
\begin{itemize}
    \item Reveal the current situation of external developers obtaining trust in top OSS communities.
    \item Propose a taxonomy of committer qualifications.
    \item Reveal the differences in expected and actual qualifications of committers.
    \item Provide practical insights for committer immigration.
\end{itemize}

%We organize the remainder of the paper as follows. Section~\ref{related_work} presents related work; Section~\ref{study_design} introduces our dataset and methodology overview. Sections~\ref{RQ1} and~\ref{RQ2} present methods and answers to the two research questions. Section~\ref{discussion} presents the discussions of the findings. Section~\ref{conclusion} concludes the paper. Section~\ref{sec: data} presents the dataset and scripts.

\section{Related Work}\label{related_work}
Many studies explore developer immigration between different roles in OSS communities, covering topics such as joining scripts, motivation, and barriers.

%Motivations for participating OSS have been extensively investigated~\cite{ye2003toward,alexander2002working,shah2006motivation,lakhani2003hackers,hertel2003motivation,von2012carrots,gerosa2021shifting,huang2021leaving}. A representative survey conducted by Von Krogh et al.~\cite{von2012carrots} reviewed the related research until 2009. They find that developers contributing to OSS are driven by intrinsic, internalized extrinsic, and extrinsic motivations. Some studies focus on the motivations in specific scenarios, including particular communities~\cite{hertel2003motivation,choi2015characteristics,spaeth2015research,bosu2019understanding}, specific developer profiles (e.g., newcomers~\cite{hannebauer2017relationship}, one-time contributors~\cite{lee2017understanding}, quasi-contributors~\cite{steinmacher2018almost}, core contributors~\cite{coelho2018we}, and students~\cite{silva2020google}), and the relation between motivation and other constructs (e.g., retention~\cite{wu2007empirical}, task effort~\cite{ke2010effects}, and participation level~\cite{meissonierm2012toward}). Researchers have revisited developers' motivations in recent years and found that motivations related to social aspects and reputation increased~\cite{gerosa2021shifting,qiu2019going,mendez2018open}.
An external developer can immigrate to different community roles by completing various activities. This process is usually called joining script in the literature~\cite{bird2007open, sinha2011entering,von2003community,ducheneaut2005socialization,trinkenreich2020hidden,ihara2014early}. The most common form of role organization in OSS communities is called the onion model~\cite{jergensen2011onion}. This model depicts roles as concentric layers with high skill, high reputation roles at the center and low technical skill and reputation at the periphery. Previous studies focus on the role transition in this model. Through a longitudinal analysis of a selection of projects in the GNOME ecosystem, Jergensen et al.~\cite{jergensen2011onion} find that prior experience in the project or the ecosystem does not seem to have a high effect on the overall centrality of a developer’s contribution. Through analyzing Eclipse projects, Sinha et al.~\cite{sinha2011entering} find that developers establish trust and credibility in a project by contributing to the project in a non-committer role, and their affiliation is another factor—although a less significant one—that influences trust. Zhou and Mockus~\cite{zhou2012make} find that the probability for newcomers to become long-term contributors is related to the capability they present through a number of tasks, the effort they devote to issue reports, and the amount of attention they receive from the project. A recent study by Trinkenreich et al.~\cite{trinkenreich2020hidden} interviewed 17 OSS developers in well-known communities and found that developers can build a career in OSS through different roles and activities with different backgrounds, including those not related to writing software. The most relevant study for our work was carried out by Bird et al.~\cite{bird2007open}. They studied the process of gaining commit rights in Apache web server, Postgres, and Python. Through survival analysis, they find that 1) the rate of committer immigration (i.e., developers who do not initially have the commit right but later become committers) is non-monotonic; 2) demonstrated technical skill and social reputation both impact the chances of becoming a committer.

Some studies investigate why developers immigrate in specific developer roles (e.g., newcomers~\cite{hannebauer2017relationship}, one-time contributors~\cite{lee2017understanding}, quasi-contributors~\cite{steinmacher2018almost}, and core contributors~\cite{coelho2018we}), and the relation between motivation and other constructs (e.g., retention~\cite{wu2007empirical}, task effort~\cite{ke2010effects}, and participation level~\cite{meissonierm2012toward}). Immigrating to a new OSS community is uneasy. Newcomers face various barriers~\cite{steinmacher2015systematic,mendez2018open,steinmacher2015social,dias2019barriers}. Besides technical barriers~\cite{lee2017understanding,hannebauer2014exploratory,avelino2019abandonment}, some non-technical barriers (e.g., communication) also hinder developers from contributing ~\cite{steinmacher2015social, tan2020first}. Because women are underrepresented in OSS projects, some studies focus on their barriers~\cite{trinkenreich2022women,qiu2019going,dias2019barriers,vasilescu2015gender}. Through a systematic review, Steinmacher et al.~\cite{steinmacher2015systematic} identified 15 different barriers in five categories: social interaction, newcomers' previous knowledge, finding a way to start, documentation, and technical hurdles. 

These studies lay a foundation for comprehending developer immigration across various roles in OSS communities. \textcolor{black}{However, the existing research particularly focusing on ``\textit{committer immigration}'' (such as Bird et al.~\cite{bird2007open}) primarily relies on interviews or limited hypotheses derived from specific projects. As a result, there is a lack of comprehensive understanding regarding the process by which external developers transition to committers, as well as a limited consideration of the influence of governance types.} Considering the critical roles of committers, we undertake a systematic analysis of community policies on committer immigration and successful immigration cases within prominent OSS communities to bridge this knowledge gap.

\vspace{-0.2cm}
\section{Study Design}\label{study_design}
We introduce the overall design of our study and the preparation and inspection of the dataset. We present the details of our methods in the later sections, along with two research questions.
\vspace{-0.2cm}
\subsection{Methodology Overview}
\begin{figure}[h]
\centering
\includegraphics[width=0.75\textwidth]{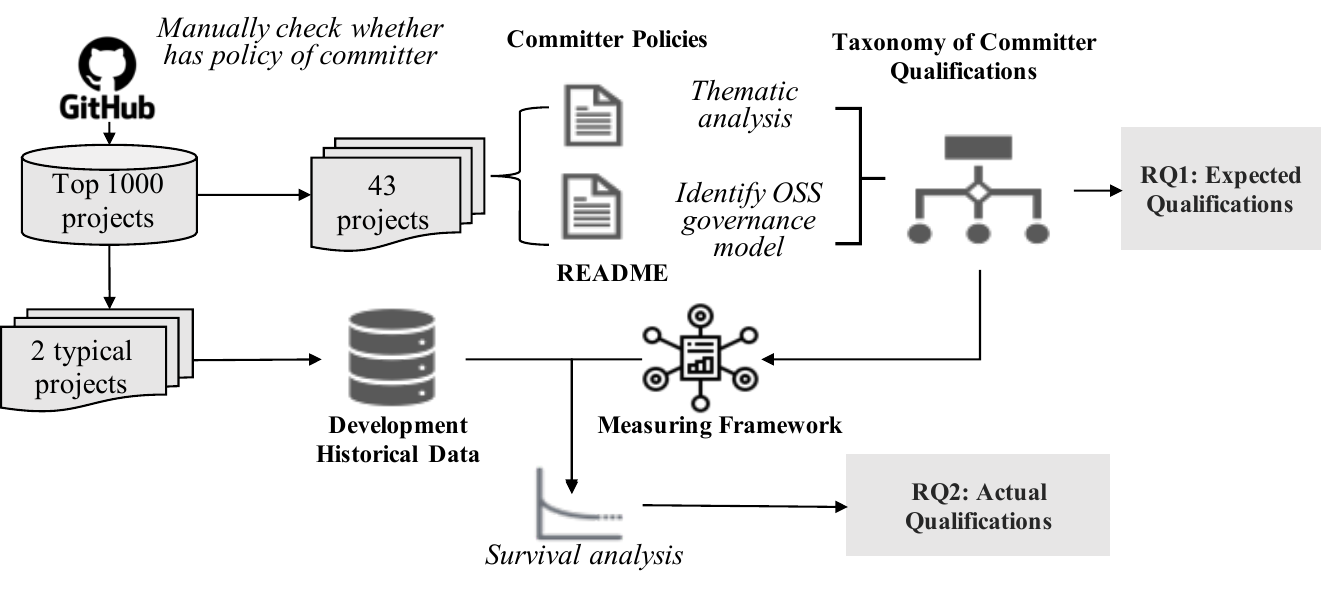}
\vspace{-0.3cm}
\caption{Overview of Research Methodology}
\vspace{-0.3cm}
\label{fig:method}
\end{figure}

Fig.~\ref{fig:method} shows the overview of our methodology. We adopted a mixed-method approach~\cite{creswell1999mixed} to conduct this study. First, we selected 1,000 of the most popular OSS projects on GitHub and conducted data inspection to motivate our study. Second, we sampled the 43 most popular GitHub projects that explicitly describe how to obtain the commit right in their communities. Then, we conducted a thematic analysis of the committer policies along with their type of OSS governance model. After this process, we constructed the taxonomy of committers' qualifications (RQ1). Then, based on the results of RQ1 and related literature, we designed a set of metrics to quantify committers' qualifications. Finally, to validate our qualitative findings and investigate the practical implementation of these qualifications, we conducted a survival analysis on two representative projects (RQ2).

\subsection{Data Preparation}\label{sec: Data Preparation}
We are interested in understanding committer qualifications in top OSS communities. Therefore, we retrieved the 1,000 most popular (i.e., with the most number of stars) GitHub repositories via the GitHub API\footnote{\url{https://docs.github.com/rest}} in Oct. 2022 as our initial dataset. Because not all these projects are suitable for our study, we adopted the following process to safeguard the dataset's quality and check each project for OSS. First, we excluded forks to avoid including the same project multiple times. Second, we excluded the projects explicitly labeled as unmaintained or not in English. Third, we carefully read the projects' descriptions and licenses to exclude non-OSS projects, e.g., education, storage, and code collection. Finally, 734 projects left. We cloned them for further analysis. 

To determine whether communities have policies on the commit right, we carefully read the README file of 734 projects. For some projects, this information exists in the GOVERNANCE file, CONTRIBUTING file, or the documents pointed to by the hyperlinks in the README file. For such cases, we also checked these documents. We found that 43 projects specify how to apply for the commit right. The characteristics of these projects are shown in Table~\ref{tab:characteristics_projects}. Most of them are popular and have large scales and long histories. These projects are used to answer RQ1.

To obtain the timestamp for the developer to obtain the commit right, we analyzed their commit logs following previous work~\cite{sinha2011entering,ihara2014early}. Git distinguishes the roles of ``author'' and ``committer''. The ``author'' is the person who originally created the content or made the changes, whereas the ``committer'' is the person who merged those changes to the main repository. So, if a developer sends a patch to a project and one of the core members applies the patch, both developers get credit~\cite{chacon2014pro}. For example, \textit{dev} submitted a pull request (PR) which was accepted and then committed by \textit{cmit} (the person who has the commit right). The name of \textit{dev}/\textit{cmit} appears in the \textit{Author}/\textit{Commit} field of the related commit. Thus, to decide whether a developer has the commit right, we checked whether this developer's name has appeared in the \textit{Commit} field of the project commit log. When a developer's name appears in the \textit{Commit} field for the first time, we consider that the developer has obtained the commit right since this time. Thus, if a developer's initial commits are committed by someone else but their name appears in the \textit{Commit} field of subsequent commits, we consider this as the signal that external developers transit to the role of a committer. 
\textcolor{black}{It is worth noting that the use of Git web interfaces such as GitHub may not ensure complete accuracy in recording committer information. GitHub supports three types of merge methods: 1) git merge --no-ff, 2) squashing merge, and 3) rebase and merge. With the exception of the first method, all the others can accurately record the identity of the committer~\cite{GitHub2024merge}. However, this approach can still be effective for the two selected OSS projects in RQ2. The reasons for this will be further discussed below.}

\begin{table}[]
\caption{Characteristics of the Projects for RQ1}
\small
\label{tab:characteristics_projects}
\begin{tabular}{rrrrrr}
\toprule
    & \#commit & \#author & \#committer & \#star & \#age (Y) \\ \midrule
min & 285       & 60     & 7   & 16,398   & 3     \\
mid & 16,360   & 884    & 180   & 35,315   & 9     \\
max & 141,277    & 6,130  & 1,602    & 144,780  & 32    \\ \bottomrule
\end{tabular}
\end{table}

To answer RQ2, we focus on the two representative OSS projects: \textit{Node.js}\footnote{\url{https://github.com/nodejs/node}} and \textit{Terraform}\footnote{\url{https://github.com/hashicorp/terraform}}. %They are on large scales and have rich cases of committer immigration, which ensures the reliability and validity of the survival analysis. 
The information on these two projects is shown in Table~\ref{tab:information_two_projects}. We chose these two projects for the following reasons. \textbf{1) They are representative projects in different fields and with different OSS governance models.} Thus, the results of RQ2 can confirm and enhance our understanding of the qualitative findings in RQ1. \textit{Node.js} is an open-source, cross-platform JavaScript runtime environment. It uses a \textit{community-driven governance model} in which a small group of core maintainers is responsible for leading and making decisions for the project. \textit{Terraform} is an infrastructure as code tool that enables developers to safely and predictably provision and manage infrastructure in any cloud. It uses a \textit{company-backed governance model}, where Hashicorp plays a leading role in its development and decision-making. \textbf{2) They have rich cases of committer immigration.} The number of cases is 125 (\textit{Node.js}) and 98 (\textit{Terraform}), respectively, ranking the top three among the projects with instructions on the commit right.\footnote{Apache project -- Flink (https://github.com/apache/flink) has the second largest number of immigration cases. We decided not to consider this project when answering RQ2 because it does not manage the issues on GitHub. Analyzing developer identities across different platforms is difficult.} These rich cases can meet the modeling requirements of survival analysis (EPV rule~\cite{nunez2011regression,vittinghoff2007relaxing}) and indicate that these projects are relatively open to external developers. \textcolor{black}{\textbf{3) Their merge methods can significantly improve the accuracy of the aforementioned approach in identifying committer immigration.} For the \textit{Node.js} community, their collaborator guide~\cite{Nodejs2024Landing} explicitly advises against using GitHub's ``Merge pull request'' button and recommends use ``curl -L https://github.com/nodejs/node/pull/xxx.patch | git am --whitespace=fix'' to merge PRs. Through this way, the commit message can accurately record the author who wrote the patch and the committer who merged the patch. Additionally, the README.md of \textit{Node.js} dynamically maintains the list of current collaborators after Nov. 2017~\cite{Nodejs2024Collaborators}, allowing us to validate the accuracy of our method by comparing it with the modification history of this list. Out of 62 cases, 57 were confirmed, indicating the acceptable accuracy of our method. For the \textit{Terraform} community, we consulted with its maintainers, who mentioned that in addition to using ``curl -L https://github.com/hashicorp/terraform/pull/xxx.patch | git am --whitespace=fix'' for merging PRs,  they also rely on GitHub's ``rebase and merge'' function to ensure accurate updates of committer information for PRs from external developers. This method is similar to the process of applying git apply email patches. After manually examining all the 98 immigration cases within the \textit{Terraform} community, including analyzing developers' profiles and community discussions, it was determined that 79 cases had accurate committer information. For 14 cases, there was uncertainty as no information regarding their immigration was found. Additionally, 5 cases were identified as being incorrect, as the date of immigration was inaccurately recorded. This indicates that our approach for identifying committer information can generally yield reliable results in both \textit{Node.js} and \textit{Terraform} communities. For the cases that we confirmed incorrect identification, we decided to exclude them when we answer RQ2.}

\begin{table}[]
\caption{Characteristics of the Projects for RQ2}
\vspace{-0.3cm}
\small
\label{tab:information_two_projects}
\begin{tabular}{rrrrrrr}
\toprule
    & \#commit & \#author & %\#committer & 
    \#immigration & \#star & \#age (Y) \\ \midrule
Node.js & 26,654       & 3,049    %& 237  
& 125 & 90,187  & 8  \\
Terraform & 30,394    & 1,923  %& 963  
& 98  & 34,071  & 9   \\ 
\bottomrule
\end{tabular}
\end{table}

%To select the OSS projects to be studied more appropriately, we take into consideration of two main factors. We prioritize the scale of committer immigration in the project. The number of committers immigration in node is 198 and 92 in flink, ranking the top two among the projects with instructions on the commit right. The large scale of committer immigration can not only met the rigid requirements of independent variable multiples in subsequent modeling, but also indicate that this project has a high degree of acceptance of external developers, which is more suitable to analyse. We also consider whether the instruction of committer right of the project has a high degree of coincidence with the variables to be quantified. The instruction in the node project is rich and basically coincides, while the instruction in the flink project is relatively less. However, there are inevitable omissions since the filter of instruction is carried out manually, so we believe that priority selection of projects with large scale of committer immigration can also reduce the impact of such omissions by analyzing actual data.

%\subsection{Data Inspection}

\subsection{Motivational Data Inspection}
We analyzed the proportion of external developers who successfully migrated to committers to motivate our study. %we conducted  from the following perspectives: 1) characteristics of the communities with/without descriptions on the commit right; 2) 

\begin{comment}
\begin{figure}[h]
\centering
\includegraphics[width=0.48\textwidth]{fig/project_char.pdf}
\caption{Comparison of Projects' Characteristics between $w/\_C\_Descr$ and $w/o\_C\_Descr$. Outliers are removed.} 
\label{fig:project_char_comparsion}
\end{figure}

As shown in Figure~\ref{fig:project_char_comparsion}, the projects that have descriptions about the commit right (denoted as ``$w/\_C\_Descr$'') are generally larger, have more developers, are more popular, and have longer histories compared with the projects that do not have descriptions about the commit right (denoted as ``$w/o\_C\_Descr$''). However, exceptions exist, such as the project: vim\footnote{\url{https://github.com/vim/vim}}. Although its development history is over 18 years, most of the code is written and maintained by its benevolent dictator --- \textit{Bram Moolenaar}~\cite{Bram2022Vim}. Therefore, it is almost impossible for external developers to obtain the commit right in this community.
\end{comment}

\begin{figure}[h]
\centering
\includegraphics[width=0.6\textwidth]{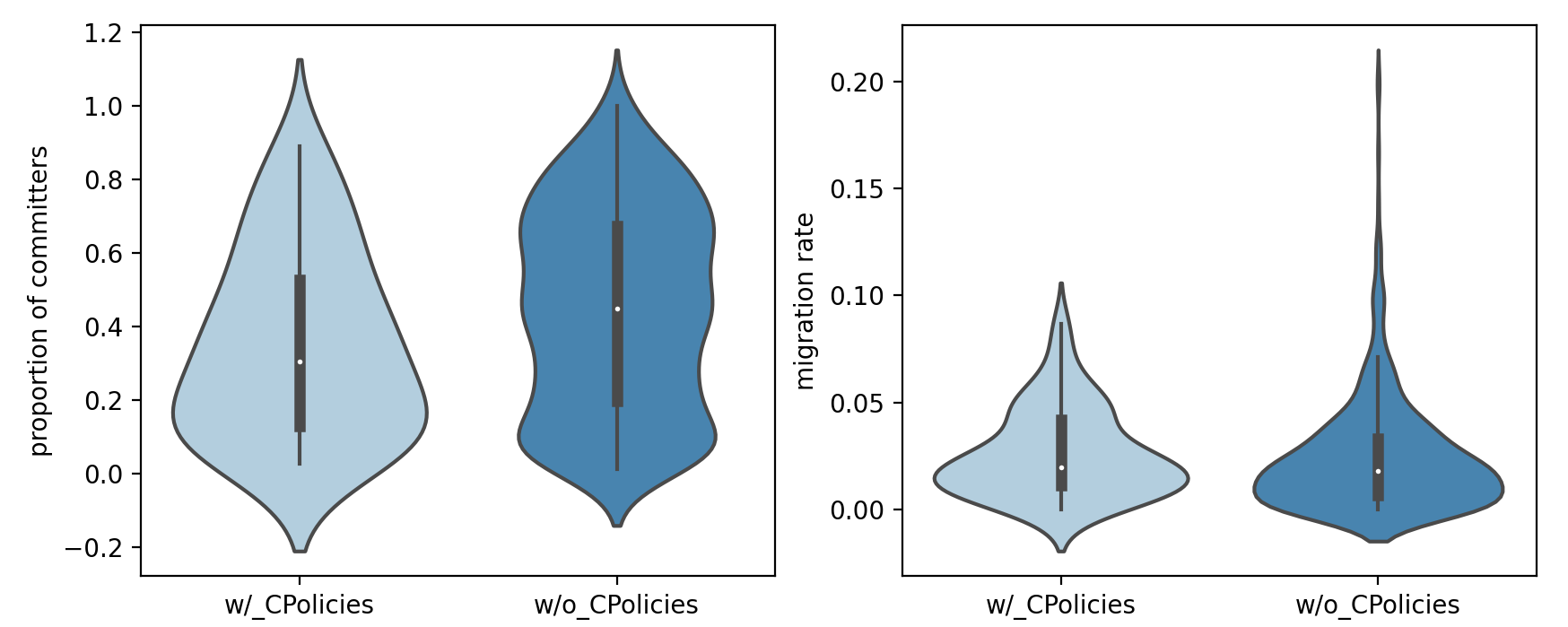}
\vspace{-0.3cm}
\caption{Committer Proportion and Immigration Rate} 
\vspace{-0.3cm}
\label{fig:project_immigration_comparsion}
\end{figure}

As shown in Fig.~\ref{fig:project_immigration_comparsion}, the proportions of committers in top OSS communities are concentrated between 0.25 and 0.5. This value in the projects with committer policies ($w/\_CPolicies$) is significantly lower than that in the projects without committer policies ($w/o\_CPolicies$) (Mann-Whitney U test~\cite{nachar2008mann}: P-value = 0.03, Cliff's delta = -0.20). It indicates that the committers are rarer in the communities with policies on the commit right. As for the committer immigration rate, the two types of communities have no significant differences. The median rates are lower than 0.02. It means that less than 2\% of developers who do not initially have the commit right can eventually become committers, so it is challenging for developers to enter the circle of trust of top OSS communities. This great challenge motivates our study and demonstrates the significance of this study.

%\gy{This situation with great challenge for external developers is the motivative exmaple of this study, which demonstrates the contribution and value of our study.}

\begin{comment}
\begin{figure}[h]
\centering
\includegraphics[width=0.45\textwidth]{fig/committer_trend.pdf}
\caption{Trend of Projects' Committer Proportion. Outliers are removed. Each box represents the distribution of the proportion of committers among developers who have ever contributed from the beginning of the project to the current year.} 
\label{fig:trend}
\end{figure}

Figure~\ref{fig:trend} shows the trend of the committer proportion of OSS communities. It indicates that the proportions of committers in the communities are gradually declining over time. This is reasonable because, in the early stage of a project, only the creators of the project usually have the commit right. As time goes by, more and more external developers join the project, but not everyone can eventually get the commit right, so the proportion of committers is gradually decreasing. It is interesting that in the early stage of the projects, the proportion of committers in $w/\_C\_Descr$ is usually higher than that in $w/o\_C\_Descr$. About four years later, this phenomenon has been reversed. It implies that welcoming external developers to apply for the commit right may draw more contributors to communities.
\end{comment}

\section{RQ1: Expected Qualifications}\label{RQ1}
\subsection{Methodology}\label{sec:RQ1_M}
To understand how OSS communities grant the commit right, we focused on the 43 projects that have policies of the commit right. In addition to the analysis of the policies of the commit right, we also identified their community governance models.

\subsubsection{Analyzing the Policies of the Commit Right.}
We reserved each policy file as a single document for further manual analysis.
We applied a thematic analysis~\cite{cruzes2011recommended} to extract the information related to committer qualifications. We randomly sampled 13 (30\%) of the 43 documents for a pilot analysis. We read and reread these documents to become familiar with them. We then conducted an open coding procedure on these documents, grouping codes into themes that were conceptually similar. This process produced a codebook that reveals different codes and themes of committer qualifications. The first three authors performed the above procedure together. 

Then, the second and third authors (named as inspectors) individually performed extended analysis. The inspectors used the above set of codes to code the remaining 30 documents for reliability analysis. Codes not in the codebook are added to a new theme named \textit{Pending}. The inter-rater reliability during the independent labeling was 0.79 (Cohen's Kappa), which indicates substantial agreement between the inspectors and demonstrates the reliability of our coding schema and procedure~\cite{hallgren2012computing}. After completing the labeling process, the two inspectors discussed their codes, and the discrepancies were discussed with the first author. For the codes classified as \textit{Pending}, the first three authors discussed together to determine whether new codes/themes need to be added. After this process, three new codes and one theme were added. Finally, we obtained 26 codes and nine themes.

\subsubsection{Analyzing the Communities' Governance Models.} Because the policies about committers are related to the communities' governance models, we identified each project's governance model considering related studies~\cite{o2007governance,o2007emergence} and the projects' README files and GOVERNANCE files. Types and corresponding quantities of different models are shown as the following. We use this information to analyze the differences in committer qualifications among different governance models.

\textbf{Community-driven Model (22 projects)}.
This model encourages broad community participation and decision-making. Decisions are reached through discussions, voting, or consensus, and involve active participation from community members. Community members contribute to the project's development through code contributions, suggesting features, reviewing issues, and more.

\textbf{Company-backed Model (12 projects).}
One or more companies play a leading role in the development and decision-making of the OSS project. Companies may provide funding, technical support, and human resources, and have significant influence over the project's direction. However, this model also requires ensuring community participation and transparency.

\textbf{Foundation/Organization Governance Model (6 projects).}
In this model, an independent foundation or organization is responsible for managing and guiding the development of the project. The foundation or organization typically consists of multiple stakeholders who establish policies, manage finances and legal matters, and promote community participation and project sustainability.

\textbf{Core Maintainer Model (3 projects).}
In this model, a small group (usually typically less than five) of core maintainers is responsible for leading and making decisions for the project or organization. Core maintainers are typically project founders or members with significant contributions. They are responsible for the project's direction, code review, version releases, and other critical decisions.

\vspace{-0.1cm}
\subsection{Results}\label{sec:RQ1_R}

We identified 26 codes and nine themes about committer qualifications. We present the codes and themes, followed by a discussion of differences in qualifications among various governance models.

\subsubsection{Taxonomy of Committer Qualifications.}
\begin{figure*}[h]
\centering
\vspace{-0.3cm}
\includegraphics[width=0.9\textwidth]{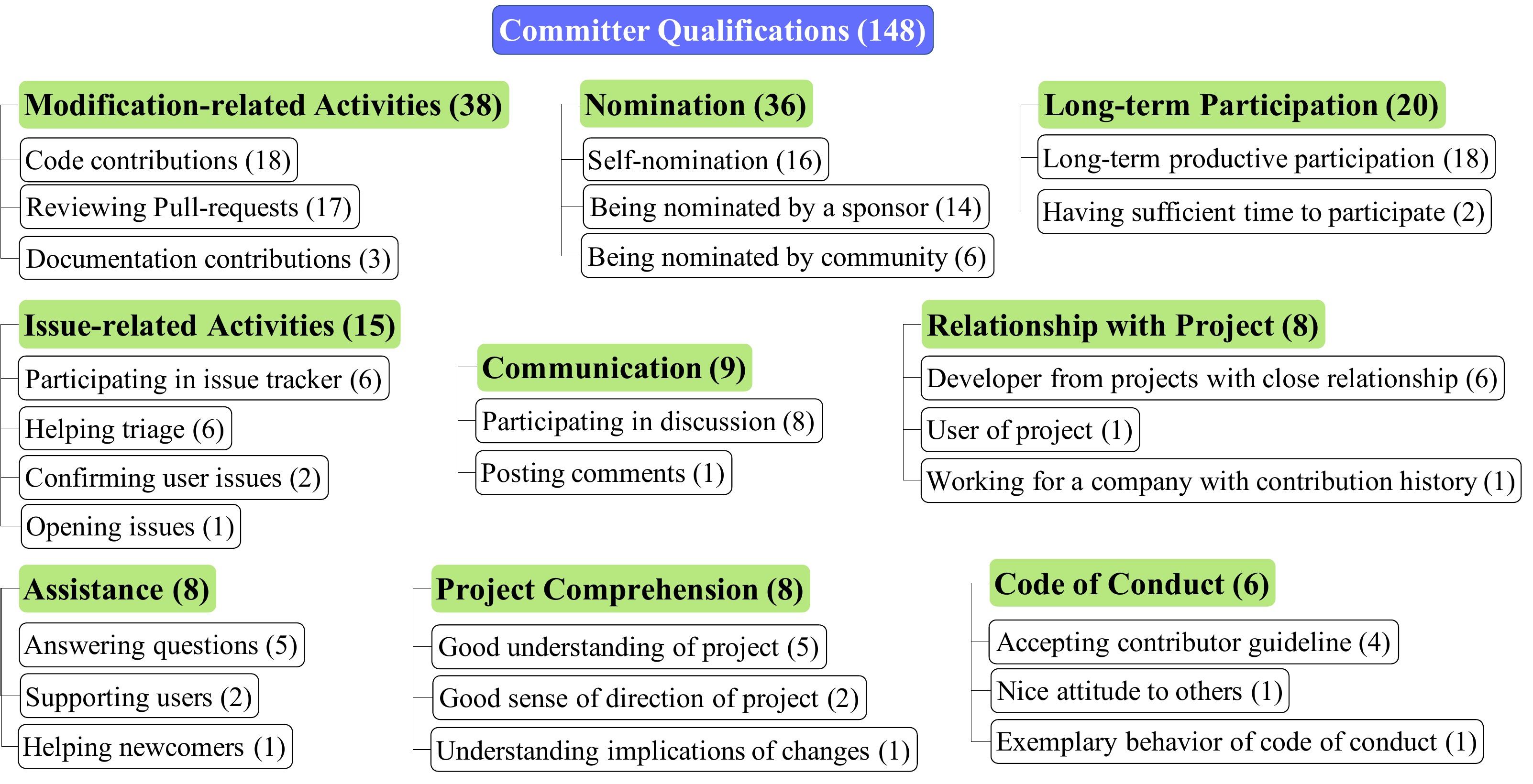}
\vspace{-0.3cm}
\caption{Taxonomy of Committer Qualifications. The number in the parentheses represents the frequency of the codes/themes.} 
\vspace{-0.3cm}
\label{fig:qualifications}
\end{figure*}

Fig.~\ref{fig:qualifications} depicts the qualifications of committers, which involves 26 codes from nine themes. To reflect communities' interest in specific codes/abilities, we show the frequency of the codes in parentheses. Then we describe and exemplify each theme in detail. During analysis, we refer to the 43 projects as P1 - P43.

\textbf{Modification-related Activities.} This theme focuses on modifying software projects, with \textbf{making code contributions} being the primary requirement to become a committer. %To be a committer, the primary requirement is \textbf{making code contributions}. 
Code is the heart of OSS projects~\cite{o1999lessons}. Contributing code is still one of the most popular ways to get into open source, and it is also the most visible way for developers to demonstrate their abilities~\cite{ye2003toward}. However, not all code contributions can win the trust of OSS communities. Generally, OSS communities have requirements for the code contributions of their expectant committers, such as quality, difficulty, and quantity. For example, P30 mentions that \textit{developers can be nominated as committers if they finish two or more tasks of medium difficulty and fix one or more tasks of hard difficulty}. \textbf{Reviewing PRs} and participating in the PR discussions at the same time is also a great way to introduce developers themselves to communities. Many projects are overloaded~\cite{tan2020scaling}. Reviewing PRs can mitigate the burden of current committers. Thus, it is an essential requirement of nominees, almost equally important as ``code contribution''. For example, P22 mentions that \textit{nominees should have provided good feedback to other contributors and filed and reviewed PRs to fix medium or high-priority bugs}. In addition to code contributions, \textbf{documentation contributions} (such as developer guides, user guides, examples, or specifications) are also mentioned by three communities. Although people usually think that documentation contributions are suitable for newcomers~\cite{tan2020first}, in fact, non-code contributions are very desired and appreciated. It is because the success of an open-source project depends on much more than simply the software~\cite{midha2012factors}.

\textbf{Nomination.} Developers who wish to become committers should first be nominated as candidate committers. There are three types of nominations. The most common one is  \textbf{self-nomination}, as stipulated by P8, ``\textit{if you would like the commit access, please send an email to the code owners list with the GitHub user name that you want to use and a list of 5 non-trivial PRs that were accepted without modifications}''. Another common way is \textbf{being nominated by a sponsor}. This sponsor is usually a developer with the commit right. For example, P19 asks that ``\textit{current maintainers may nominate a contributor and confer maintainer status}''. The third way is \textbf{being nominated by the community}. Generally, this way is not proactive. As P5 mentions, ``\textit{when the time comes, Microsoft will reach out and help make you a formal team member}''. However, please note that nomination just means that developers become candidate committers instead of obtaining the commit right.

\textbf{Long-term Participation.} Being granted the commit right implies gaining the trust of communities~\cite{gharehyazie2015developer}. It requires a significant amount of time devoted to engagement in OSS projects, which has two meanings. First, it means \textbf{long-term productive participation}. As P36 mentions, ``\textit{there's someone who has been making consistently high-quality contributions to Homebrew and shown themselves able to make slightly more advanced contributions}''. P17 also has a similar stipulation --- ``\textit{the contributor has opened and successfully run medium to large PRs in the past 6 months}''. Second, it means \textbf{having sufficient time to participate}. For example, P43 mentions that ``\textit{becoming a maintainer means that you are going to be spending substantial time (>25\%)}''.

\textbf{Issue-related Activities.} OSS communities manage bug reports and feature requests via tools called ``issue trackers''. \textbf{Participating in the issue tracker} is an important way to get involved in OSS communities~\cite{jergensen2011onion}, and it is also a requirement for obtaining the commit right. Specifically, the following activities can increase the opportunity of being granted the commit right. First, developers can \textbf{help bug triage} and \textbf{confirm user issues}. Because bugs are often poorly reported, diagnosing and triaging a bug can help developers save time with the legwork of figuring out the specifics of the problem~\cite{bettenburg2008makes}. As P10 mentions, ``\textit{nominees should triage and confirm user issues}''. P1 also says that the contribution of \textbf{opening issues} is one of the criteria for evaluating nominees.

\textbf{Communication.} Contributing to OSS communities involves working with others collaboratively, so effective communication is key~\cite{guzzi2013communication}. Therefore, communication is also a criterion for evaluating nominees. On the one hand, potential committers can express their desire to enter the circle of trust by actively \textbf{participating in community discussions}. For example, P17 mentions that ``\textit{the contributor is active on Traefik Community forums or other technical forums/boards such as K8S slack, Reddit, Stack Overflow, hacker news}''. \textbf{Posting comments} on others' PRs/issues is also one of the criteria for evaluating nominees, which is mentioned by P1.

\textbf{Relationship with Project.} Some communities are willing to grant the commit right to developers with specific identities. Among different identities, the \textbf{developers from projects with close relationships} is the most frequently highlighted category, e.g., projects in the same organization or belonging to their upstream/downstream. Because these projects are often in the same domain, their developers tend to have relevant experience and are, therefore, likely to be trusted by the community. P14 mentions that they are willing to grant the commit right to \textbf{users} of Jekyll. More and more corporations devote employees and company resources to OSS projects to achieve their commercial goals~\cite{zhang2019companies}. We notice that some communities clearly indicate that they would grant the commit right to \textbf{employees of companies with contribution history}. For example, P2 states that ``\textit{employed by a company with a history of contributing to Flutter}''.

\textbf{Assistance.} In OSS communities, collaboration among developers from diverse backgrounds is crucial for success~\cite{vasilescu2015gender}. One of the key factors in gaining the commit right is the willingness to offer assistance in various forms. This assistance includes \textbf{answering questions}, \textbf{supporting users}, and \textbf{helping newcomers} integrate into the community. The nomination process takes into account the level of help provided to end-users and novice contributors. This means that developers who actively engage in addressing user inquiries, providing support, and guiding newcomers will be recognized and considered for commit rights. By actively participating in these support activities, developers not only contribute to the growth and sustainability of the community but also demonstrate their commitment to fostering a welcoming and inclusive environment. This recognition of their efforts can lead to increased trust and responsibility within OSS communities.

\textbf{Project Comprehension.} OSS projects with mature management of the commit right are usually complex and large, involving many technical difficulties and having intricate logical structures. If developers do not understand the whole system sufficiently and are blindly granted the commit right, the software may have low quality. Therefore, \textbf{a good understanding of projects} is a criterion for being granted the commit right. Besides having a good understanding of projects' current status, nominees should also have \textbf{a good sense of the direction of projects} and \textbf{understand the implications of changes}. This ensures that changes will not have negative effects, e.g., system crashes, and that projects are always moving in the right direction. For example, P39 stipulates that ``\textit{committers should be familiar with the codebase, and enough context to understand the implications of various changes and a good sense of the will and expectations of the project team}''.

\textbf{Code of Conduct.} To decrease harmful actions, OSS projects use codes of conduct to promote ethical behavior~\cite{tourani2017code}. \textbf{Accepting contributor guidelines} is an inexorable requirement for any OSS developers, certainly including potential committers. Not only that, OSS communities may have higher expectations for nominees. For example, they hope that nominees can show \textbf{exemplary behavior of the code of conduct} and have \textbf{a nice attitude to others} because committers frequently interact with other developers, e.g., reviewing code, answering questions, and posting comments. For example, P2 mentions that nominees should ``\textit{demonstrate exemplary behavior that closely aligns with our code of conduct}''.

\subsubsection{Comparsion of Committer Qualifications Among Different OSS Governance Models.}
\begin{figure*}[h]
\centering
\vspace{-0.4cm}
\includegraphics[width=0.95\textwidth]{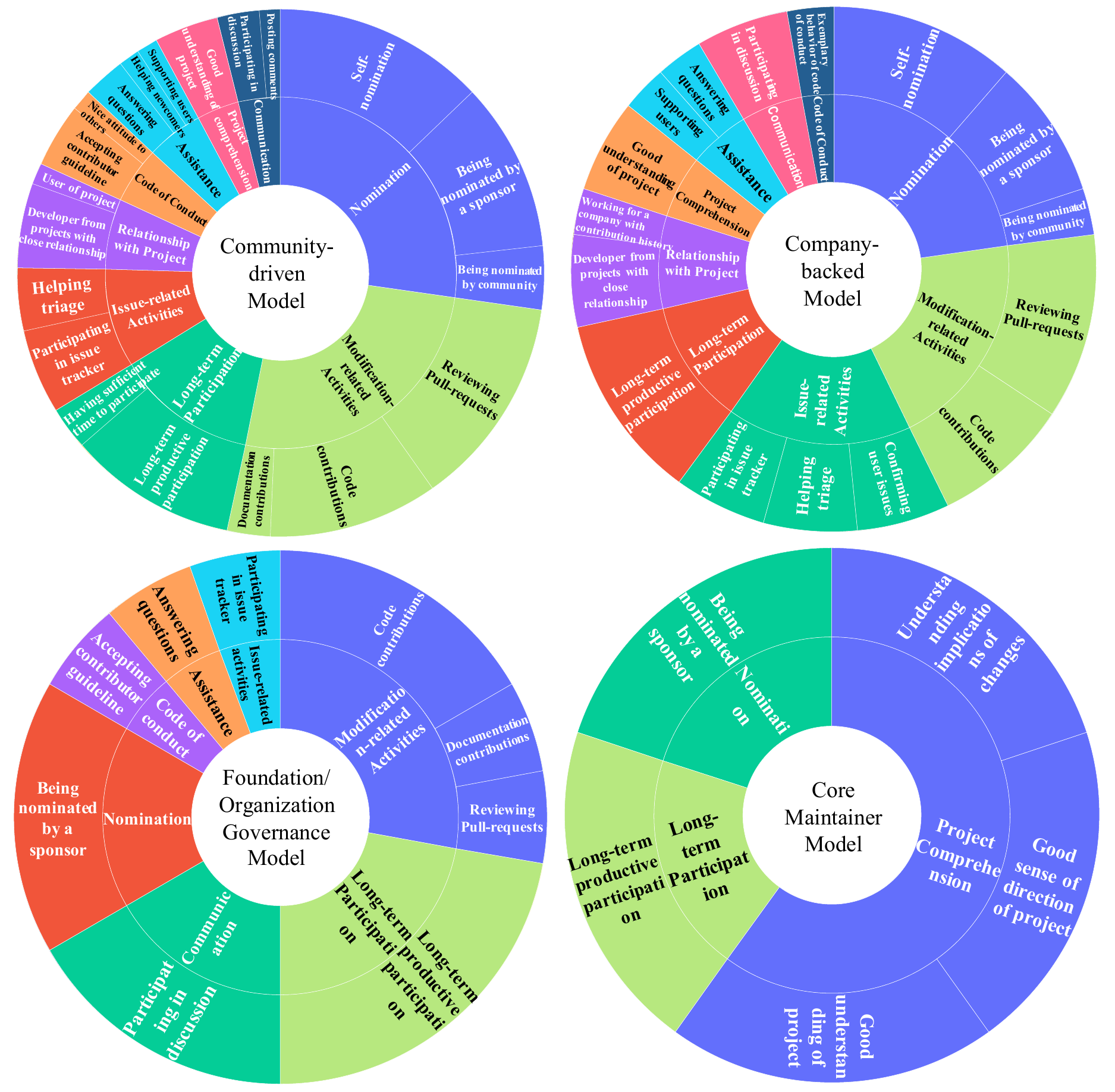}
\vspace{-0.3cm}
\caption{Qualifications of Committers along with Different OSS Governance Models. \textcolor{black}{The sunburst chart illustrates the hierarchical relationship between themes and codes, as well as the proportional representation of different codes. The innermost ring represents the themes, while the outer rings depict the codes associated with each theme. The size of each segment corresponds to the proportion of codes within a theme. The different colors further distinguish different themes.}} 
\vspace{-0.5cm}
\label{fig:qualifications_4models}
\end{figure*}

Fig.~\ref{fig:qualifications_4models} consists of four sunbursts, each representing the qualifications of committers in a type of OSS governance model. The four governance models have different emphases on committer qualifications. 

The \textbf{community-driven model}, characterized by its openness and transparency, sets the rigorous standards for individuals aspiring to become committers. This is partly due to the large number of projects (almost half of the projects examined) following this model. In these projects, the evaluation of potential committers focuses on their involvement in various modification-related activities, including reviewing PRs, making code contributions, and contributing to documentation. Long-term participation and engagement in issue-related activities, as well as offering assistance to others, are also key factors considered in the nomination process. The nomination process within this model is flexible, allowing potential committers to be put forward by various entities, including self-nomination, sponsorship from existing committers or project leaders, or recognition from the broader project community.  

In the \textbf{company-backed model}, although the qualifications for becoming a committer share similarities with the community-driven model, there are notable differences. One key distinction lies in the emphasis placed on issue-related activities, particularly in terms of confirming and addressing user issues. This stems from the goal of this model, which is to provide a reliable and user-centric experience~\cite{o2007emergence}. Another key distinction is that developers who come from companies with a contribution history to the project are often seen as reliable contributors, and these developers may be more likely to become committers. This is because that these developers are driven by the companies' commercial goals, so they are generally active and responsible.

In the \textbf{foundation/ organization governance model}, in addition to modification-related activities, there is a greater focus on developers' long-term participation and communication. Unlike in some other models, developers seeking to commit rights in this model typically need to be nominated by a sponsor rather than being able to self-nominate. These requirements encourage developers to actively engage with the existing contributors, gain their trust/ endorsement and support, and demonstrate a long-term commitment to the project's goals and values.

Three projects follow the \textbf{core maintainer model}, which places a strong emphasis on project comprehension. This aspect receives relatively less attention in other models. Project comprehension requires developers to possess an in-depth understanding of the modifications' implications and projects' direction. In such projects, developers are expected to invest significant effort in comprehending the intricacies of the codebase, architecture, and overall project structure. This understanding is crucial for making informed decisions, implementing changes effectively, and ensuring the project's long-term stability and success.
Moreover, this model also emphasizes long-term and productive participation from developers. It expects them to actively engage with the project community, contribute consistently, and take on responsibilities beyond mere code modifications.
The combination of project comprehension and long-term participation places high demands on developers in the core maintainer model. It requires them to have not only strong technical skills but also a deep understanding of the project's goals, context, and future directions.

\vspace{1mm} 
\begin{mdframed}[linecolor=gray,roundcorner=12pt,backgroundcolor=gray!15,linewidth=3pt,innerleftmargin=2pt, leftmargin=0cm,rightmargin=0cm,topline=false,bottomline=false,rightline = false]
 %, skipabove=10pt, skipbelow=10pt
  \textbf{Summary for RQ1:} We identify 26 codes and nine themes related to committer qualifications. The importance of these qualifications varies, with the most emphasized themes being modification-related activities, nomination, long-term participation, and issue-related activities. Communication, relationship with the project, assistance, project comprehension, and code of conduct are also important, but their frequency of mention decreases in that order. Different governance models emphasize certain qualifications, such as \textit{community-driven models} with the most abundant assessment dimensions, \textit{company-backed models} valuing contributions from affiliated developers and resolving user issues, \textit{foundation/organization models} focusing on long-term commitment and communication, and \textit{core maintainer models} emphasizing project comprehension and long-term participation. These findings reveal the diverse criteria for committer selection in OSS projects.
\end{mdframed}
\vspace{1mm}

\section{RQ2: Actual Qualifications}\label{RQ2}

\subsection{Methodology}\label{sec:RQ2_M}
To investigate how projects select committers in practice, we conducted a quantitative analysis of Node.js and Terraform. The reason for selecting these two projects and the methods for identifying the committer immigration are explained in Section~\ref{sec: Data Preparation}. First, we defined eight sets of metrics to quantify the committer qualifications, and then we conducted a survival analysis to evaluate the significance of different types of qualifications for becoming committers.

\subsubsection{Defining Metrics.}
In Section~\ref{sec:RQ1_R}, we conducted an analysis to identify and characterize committer qualifications. Among the nine themes identified, one theme stood out as being different from the others. The theme of ``nomination'' does not directly reflect developers' experiences and efforts. As a result, we focused our quantitative analysis on the other eight themes that are more closely related to developers' qualifications. Table~\ref{tab:metrics} shows the definitions of these metrics. These metrics are proposed based on the taxonomy of committer qualifications.\footnote{It should be noted that not all the codes have corresponding metrics because 1) the relevant data is not recorded in the software development supporting tools, e.g., ``\textit{supporting users}''; 2) some codes are relatively abstract and difficult to be quantified, e.g., ``\textit{accepting contributor guideline}''.} Each metric is specific for a developer over a period of time.
They are widely used in previous research~\cite{zhou2012make,bird2007open,mockus2002two}. Most metrics can be directly calculated through GitHub API~\cite{GitHub2022API}. For the sake of brevity, we only discuss the metrics that are difficult to obtain through GitHub API directly. 

\begin{itemize}
    \item \textbf{M8: $\#communicator$.} This metric calculates the number of developers that a developer communicates with. We focused on communication that is related to issues, PRs, and commits. If two developers left comments on the same issue, PR, or commit, we treated both as communicators of each other. 
    \item \textbf{M9: $from\_company$.} To determine whether a developer is supported by a company, we analyzed the domain name of the email they used to submit the code and also conducted manual confirmation (e.g., we filtered out information such as ``university''). 
    \item \textbf{M14: $\#comment_{newcomer}$.} To determine whether a developer has assisted others, we focused on their communication behavior with newcomers. Many OSS communities tag the issues suitable for newcomers with labels such as ``good first issue''~\cite{tan2020first}. Thus, we calculated the number of comments posted by a developer under this type of issue. 
    \item \textbf{M16: $\#issue_{new\_feature}$.} To be a committer, developers should have a good sense of the direction of projects, which can be reflected by the new features proposed. Therefore, we calculated the number of issues related to new features opened by a developer. 
    \item \textbf{M18: $\#comment_{offensive}$.} Having a nice attitude toward others is one of the qualifications of committers. Therefore, we counted the number of offensive PR/issue/commit comments a developer posts. To determine whether a comment is offensive, we used the \textit{profanity-check} package, which is a fast and robust Python library to check for offensive language in strings, using an SVM model trained on 200k human-labeled samples of clean and profane text strings.\footnote{\url{https://github.com/vzhou842/profanity-check}} \textcolor{black}{Despite the tool being named ``\textit{profanity-check}'', it is capable of detecting offensive language, including comments that are rude, disrespectful, or likely to make someone leave a discussion, as well as profanity.}
\end{itemize}

\begin{table*}[]
\scriptsize
\captionsetup{justification=centering}
\caption{Metrics for Quantification of Committer Qualifications.}
\vspace{-0.3cm}
\renewcommand\arraystretch{1}
\label{tab:metrics}
\begin{tabular}{p{3cm}<{\centering}|p{4cm}<{\centering}|p{5.5cm}}
\toprule
\textbf{Theme Obtained in RQ1}  & \textbf{Metric} & \multicolumn{1}{c}{\textbf{Definition}} \\ \midrule
\multirow{3}{*}{Modification-related Activities} 
& $\#PR_{open}$ (M1) & $\bullet$ number of PRs opened\\
& $\#PR_{review}$ (M2) & $\bullet$ number of PRs reviewed\\ 
& $\#commit$ (M3) & $\bullet$ number of commits authored \\\hline
\multirow{2}{*}{Long-term Participation} & \multirow{2}{*}{$\#days_{active}$ (M4)} & $\bullet$ number of days the developer has activities, e.g., contribute code or left comment\\ \hline
\multirow{2}{*}{Issue-related Activities} & $\#issue_{open}$ (M5) & $\bullet$ number of issues opened \\
& $\#issue_{triage}$ (M6) & $\bullet$ number of issues triaged \\ \hline
\multirow{4}{*}{Communication} & $\#all\_comment = \#comment_{PR} + \#comment_{issue} + \#comment_{commit}$ (M7) & $\bullet$ number of comments posted including PR comments, issue comments, and commit comments\\
& $\#communicator$ (M8) &  $\bullet$ number of developers communicate with\\ \hline
\multirow{10}{*}{Relationship with Project} & $from\_company$ (M9) & $\bullet$ whether the developer is employed by a company\\
& $\#issue_{org}$ (M10) & $\bullet$ number of issues opened in the same organization of the project\\
& $\#issue\_comment_{org}$ (M11) & $\bullet$ number of issue comments posted under the issues opened in the same organization of the project\\
& $\#commit_{org}$ (M12) & $\bullet$ number of commits submitted in the same organization of the project\\
& $\#commit\_comment_{org}$ (M13) & $\bullet$ number of commit comments posted under the commits submitted in the same organization of the project\\\hline
\multirow{2}{*}{Assistance} & \multirow{2}{*}{$\#comment_{newcomer}$  (M14)} & $\bullet$ number of comments posted under the issues that are suitable for newcomers \\ \hline
\multirow{3}{*}{Project Comprehension} & $\#file_{modified}$ (M15) & $\bullet$ number of files modified \\
& $\#issue_{new\_feature}$ (M16) & $\bullet$ number of issues about new features opened \\
& $merge\_ratio = \frac{\#PR_{merge}}{\#PR_{open}}$ (M17) & $\bullet$ proportion of merged PRs in opened PRs \\ \hline
Code of Conduct & $\#comment_{offensive}$ (M18) & $\bullet$ whether the developer has offensive comments \\ \bottomrule
\end{tabular}
\end{table*}

\subsubsection{Conducting a Survival Analysis.} In this section, we introduce the background of survival analysis and data processing procedure.  

\textbf{Background of Survival Analysis.} After defining metrics, we can quantify committer qualifications. To determine which qualifications are significant, we applied survival analysis (also called ``Hazard rate analysis'')~\cite{miller2011survival}. This approach is used to study time-to-event data. Such data describe the length of time from a time origin to an endpoint of interest~\cite{KARTSONAKI2016263}. For example, studying time after cancer treatment until death or studying time from the manufacture of a component to component failure. Using statistical models, researchers can estimate the influence of time and other predictors on the occurrence of expected events, e.g., mortality, employment durations, business failures, etc~\cite{bird2007open}. In this study, we investigate committer immigration, i.e., we model the duration from a developer's first appearance in the community (considering different activities, e.g., commits, comments, PRs, etc.) to the time the first commit, if any, is committed by that individual. Therefore, the occurrence of the event refers to the time when a developer becomes a committer.

To characterize the rate at which events of interest occur, survival analysis uses the hazard rate function. This function can model the rate of event dependence on time and the other predictor variables. Because in our setting, most variables (e.g., $\#PR_{open}$) change over time, we decided to construct a Cox proportional hazards model with time-dependent covariates~\cite{therneau2017using}. 
The form of the hazard rate function is as shown in Equation~\ref{equ1}: 
\begin{equation}
    h(t,X)=h_{0}(t)exp(\beta_{1}X_{1}+\beta_{2}X_{2}+...+\beta_{m}X_{m})
    \label{equ1}
\end{equation}
where $h_{0}(t)$ represents the benchmark hazard rate, X is a vector of predictors, and $\beta_{i}$ is the partial regression coefficient of a predictor. %This function has two strict requirements: 1) the effects of risk factors do not change over time; 2) the relationship between the log hazard and each covariate is linear. 

In our setting, most risk factors (e.g., $\#issue_{open}$) change over time. Moreover, Bird et al.~\cite{bird2007open} find that the hazard rate of committer immigration is not linear. For this case, the easiest model to use is the \textit{piecewise constant exponential hazard rate model}~\cite{colvert1976estimation}. We assume that the purely time-dependent part $h_{0}(t)$ is fixed over each time interval. Thus,
\begin{equation}
    h_{0}(t)=exp(a_{tp_{k}}) \ for \  t\ \epsilon\ tp_{k}, where \ 
tp_{k} = (c_{k-1}, c_{k}]
\end{equation}
the intervals $(c_{k-1}, c_{k}]$ are chosen to cover the duration of the available data, and $a_{tp_{k}}$ are constants and each corresponds to a certain interval. Through piecewise function, we can flexibly check if the data supports the hypothesis that these rates change non-monotonically. Therefore, the final form of hazard rate is:
\begin{equation}
    h(t,X)=exp(a_{tp_{k}})*exp(\beta X)
\end{equation}
We used the survival package in R language to build this model~\cite{therneau2015package}.

\textbf{Data Processing Procedure.} We focused on two projects: \textit{Node.js} and \textit{Terraform}, to conduct survival analysis by building two models. We gathered their development history data through GitHub API. For each developer, the \textit{transition interval} is the time between their first appearance in the community and their first commit to a file. For censored data (i.e., the developers who do not become committers during our observation)\footnote{We included censored data for the reason that censored data reflected the reality of many studies without complete information on the time to event for all subjects. Including censored data in survival analysis is crucial/necessary for providing accurate estimates of survival probabilities.}, the interval is the time between their first appearance in the community and the time of data collection, i.e., Oct. 2022. We treated this interval as the ``response variable'' of our model, which can help us understand which factors are significant to the length of time needed for committer immigration. For each potential immigrant (i.e., developers without the commit right initially), we focused on their development activities during their transition interval. 

The predictor variables are the metrics in Table~\ref{tab:metrics}. We also considered $\#developer$ (number of developers in the community) and $\#age (Y)$ (the age (year) of the project ) as control variables.
All the variables were gathered monthly for the complete population of potential immigrants. Each piece of input data represented the variables of developer $d$ in the $ith$ month after the initial activity. For continuous variables, we calculated the cumulative value of the previous months (e.g., for $\#commit$, we counted the number of all the commits of developer $d$ prior to month $i$). For categorical variables, we considered the status of the metrics in month $i$. The response variable Y represents whether developer $d$ becomes committer in the $ith$ month (Y=1: become committer; Y=0 otherwise). Therefore, the model can help us understand which factors are significant to the time needed for committer immigration. The basic information of the final data is shown in Table~\ref{tab:regression_data}.

\begin{table}[]
\caption{Basic information on Regression Data}
\vspace{-0.3cm}
\small
\label{tab:regression_data}
\begin{tabular}{rrrrr}
\toprule
    & \#record & \#candidate & \#immigration & interval (M) \\ \midrule
Node.js & 9,228 & 2,666  & 125  & 5\\
Terraform & 7,981 & 975  & 98 & 10\\ \bottomrule
\end{tabular}\\
\footnotesize{``interval (M)'' represents the median length of time (months) needed for committer immigration.}
\vspace{-0.3cm}
\end{table}

Before constructing the Cox proportional hazards model, we investigated the distribution of the numeric variables and removed outliers by applying the method described by Z-Score~\cite{ghosh2012outliers}. We also applied the variance inflation factor (VIF) to detect multicollinearity problems for the reliability and stability of the fitted model~\cite{alin2010multicollinearity}. We found correlations between some variables in both projects, and we kept only one.
%For \textit{Node.js}, $\#communicator$ (M8) is correlated with $\#issue_{open}$ (M5), and we decided to keep M5; $\#days_{active}$ (M4) and $\#all\_{comment}$ (M7) are correlated with $\#PR_{open}$ (M1), and we decided to keep M1;$\#PR_{open}$ (M1) and $\#issue_{open}$ (M5) are correlated with $\#commit$ (M3), and we decided to keep M3. For \textit{Terraform}, we noticed that $\#all\_{comment}$ (M7) is correlated with $\#issue_{open}$ (M5), and we decided to keep M5; $\#days_{active}$ (M4), $\#issue_{open}$ (M5), $\#file_{modified}$ (M15), and $\#issue_{new\_feature}$ (M16) are correlated with $\#PR_{open}$ (M1), and we decided to keep M1. 
As for $\#commit\_comment_{org}$ (M13) and $\#comment_{newcomer}$ (M14), they are very sparse, so we deleted them when modeling.

\subsection{Results}\label{sec:RQ2_R}
\begin{figure}[ht]   
\centering
\vspace{-0.3cm}
\subfloat[Node.js]{\includegraphics[width=0.3\textwidth, keepaspectratio]{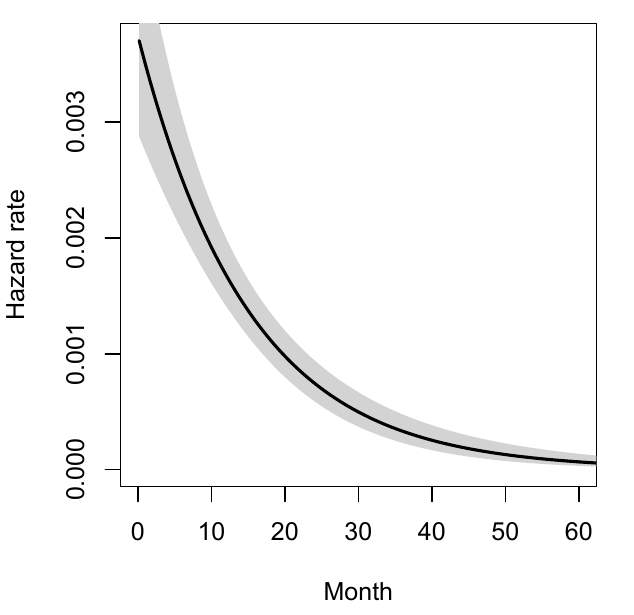}\label{fig:subfig5}}
\subfloat[Terraform]{\includegraphics[width=0.3\textwidth, keepaspectratio]{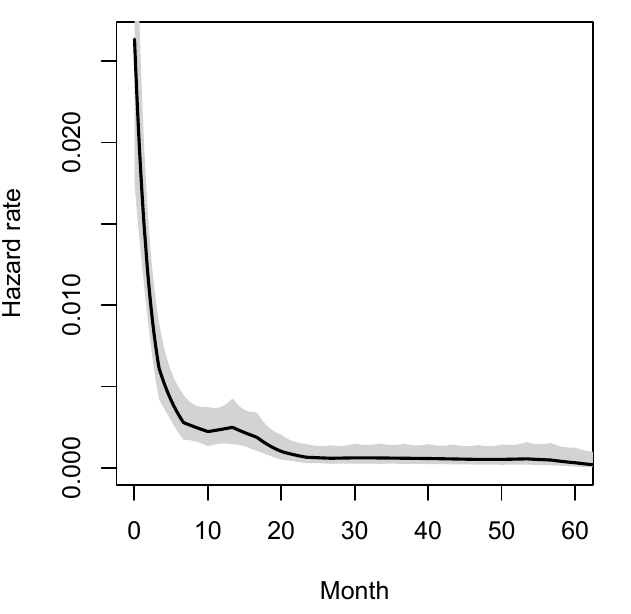}\label{fig:subfig6}}
\vspace{-0.3cm}
\caption{Smoothed Hazard Estimate}
\vspace{-0.3cm}
\label{fig:hazard_ratio}
\end{figure}

To understand the changes in the possibility of being committers, we draw the smoothed plot of the hazard rate, i.e., the chances of developers becoming committers along with time, as shown in Fig.~\ref{fig:hazard_ratio}. The hazard rates for both projects are decreasing, showing that developers find it harder to obtain commit rights as time passes. The hazard rate for \textit{Node.js} shows a gradual downward trend, while that for \textit{Terraform} drops sharply in the first ten months. \textcolor{black}{This finding differs from the non-monotonic rate of immigration discovered by Bird et al.~\cite{bird2007open}. It suggests that in modern top OSS communities, developers are required to possess a strong technical background to quickly establish trust upon entering the community. Although as time goes by, some developers have improved their capabilities and eventually joined the trust circle, it becomes more challenging to gain community trust as the participation time increases. This observation highlights the significance of technical competence in gaining recognition and trust in modern top OSS communities.}

\begin{table*}[]
\setlength\tabcolsep{1pt}
\footnotesize
\caption{Results of Cox Proportional Hazards Model Fit}
\vspace{-0.3cm}
\label{tab:model_fit}
\begin{tabular}{L{3.5cm}R{1cm}R{1cm}R{1cm}R{1cm}R{0.5cm}R{1cm}R{1cm}R{1cm}R{1cm}}
\toprule
& \multicolumn{4}{c}{\textit{Node.js} Model} && \multicolumn{4}{c}{\textit{Terraform} Model}  \\ 
\cline{2-5} \cline{7-10}
& \multicolumn{1}{c}{\textit{Coef}} & \multicolumn{1}{c}{\textit{EXP (Coef)}}    & \multicolumn{1}{c}{\textit{SE (Coef)}}    & \multicolumn{1}{c}{\textit{Z}}      && \multicolumn{1}{c}{\textit{Coef}}  & \multicolumn{1}{c}{\textit{EXP (Coef)}}   & \multicolumn{1}{c}{\textit{SE (Coef)}}    & \multicolumn{1}{c}{\textit{Z}}   \\
\midrule
\textbf{Independent Variable} &&&&&&&\\
 $\#PR_{open}$ (M1) & /\phantom{***} & / & / & / & &-0.09\phantom{***} & 0.92 & 0.14 & -0.60\\
 $\#PR_{review}$ (M2) & 0.17*** & 1.18 & 0.02 & 7.21 && 0.25*** & 1.29 & 0.06 & 4.27\\
 $\#commit$ (M3) & 0.27*** & 1.31 & 0.03 & 8.68 && 0.33*** & 1.39 & 0.03 & 10.40\\
 $\#issue_{triage}$ (M6) & -0.05*\phantom{**} & 0.95 & 0.02 & -2.06 && -0.10\phantom{***} & 0.90 & 0.07 & -1.45\\
 $\#communicator$ (M8) & /\phantom{***} & / & / & / && -0.28\phantom{***} & 0.75 & 0.20 & -1.45\\
 $from\_company$ (M9) & -0.08\phantom{***} & 0.92 & 0.10 & -0.75 && 0.53*** & 1.70 & 0.12 & 4.31\\
 $\#issue_{org}$ (M10) & -0.03\phantom{***} & 0.97 & 0.14 & -0.23 && 0.38*** & 1.46 & 0.07 & 5.10\\
 $\#issue\_comment_{org}$ (M11) & 0.24\phantom{***} & 1.27 & 0.12 & 1.93 && 0.21**\phantom{*} & 1.23 & 0.08 & 2.61\\
 $\#commit_{org}$ (M12) & 0.28**\phantom{*} & 1.32 & 0.09 & 2.95 && 0.08\phantom{***} & 1.08 & 0.06 & 1.24\\
 $\#commit\_comment_{org}$ (M13) & 0.16*** & 1.18 & 0.04 & 4.33 && /\phantom{***} & / & / & /\\
 $\#comment_{newcomer}$ (M14) & -0.05\phantom{***} & 0.95 & 0.03 & -1.71 && /\phantom{***} & / & / & /\\
 $\#file_{modified}$ (M15) & -0.41*\phantom{**} & 0.66 & 0.19 & -2.17 && /\phantom{***} & / & / & /\\
 $\#issue_{new\_feature}$ (M16) & 0.07\phantom{***} & 1.07 & 0.04 & 1.85 && /\phantom{***} & / & / & /\\
 $merge\_ratio$ (M17) & 0.54*** & 1.71 & 0.08 & 6.37 && 0.64*** & 1.89 & 0.13 & 4.85\\
 $\#comment_{offensive}$ (M18) & 0.13*** & 1.14 & 0.02 & 8.45 && 0.03\phantom{***} & 1.03 & 0.03 & 1.15\\
 \textbf{Control Variable} &&&&&&&\\
 $\#developer$ & -1.45*** & 0.23 & 0.44 & -3.33 && -1.68*** & 0.19 & 0.32 & -5.23\\
 $\#age (Y)$ & -0.30\phantom{***} & 0.74 & 0.48 & -0.63 && -0.31\phantom{***} & 0.73 & 0.40 & -0.77\\
\midrule
Likelihood ratio test & \multicolumn{4}{c}{570.2  on 15 df, p=<2e-16} && \multicolumn{4}{c}{351.3  on 13 df,   p=<2e-16} \\
Wald test & \multicolumn{4}{c}{606.8  on 15 df,   p=<2e-16} && \multicolumn{4}{c}{395.3  on 13 df,   p=<2e-16}\\
Score (logrank) test & \multicolumn{4}{c}{2860 on 15 df,  p=<2e-16} && \multicolumn{4}{c}{1078  on 13 df,   p=<2e-16}\\
%Num. obs.  & \multicolumn{4}{c}{9,228}       && \multicolumn{4}{c}{7,981} \\
\bottomrule
\multicolumn{7}{l}{***$p < 0.001$, **$p < 0.01$, *$p < 0.05$}\\
\vspace{-0.4cm}
\end{tabular}
\end{table*}

Table~\ref{tab:model_fit} shows the results of the model fit. The p-values for all three overall tests (Likelihood ratio test, Wald test, and Score test) are significant, indicating that the models are statistically significant. We describe how the factors jointly impact committer immigration. The factors $merge\_ratio$ (M17), $\#commit$ (M3), and $\#PR_{review}$ (M2) are all positively correlated with the possibility of committer immigration in both two projects. Among them, $merge\_ratio$ (M17) shows the most vital relationship. Holding the other covariates constant, an addition of $merge\_ratio$ increases the hazard (i.e., the possibility of committer immigration) by a factor of $EXP(Coef)$ = 1.71, or 71\%. We conclude that a higher value of $merge\_ratio$ is associated with a higher chance of committer immigration.
Similarly, holding the other covariates constant, an addition of $\#commit$ increases the hazard by a factor of $EXP(Coef)$ = 1.31, or 31\%, and an addition of $\#PR_{review}$ increases the hazard by a factor of $EXP(Coef)$ = 1.18, or 18\%. These results confirm our qualitative findings, i.e., developers who make more high-quality contributions and actively participate in code review are more likely to enter the community trust circle. \textcolor{black}{This finding is also consistent with the findings of Bird et al.~\cite{bird2007open} in the \textit{Apache} and \textit{Python} communities, i.e., prior history of patch submission has a very strong effect.}

For \textit{Node.js}, $\#commit_{org}$ (M12) and $\#commit\_comment_{org}$ (M13) have positive relationships with the possibility of committer immigration. It indicates that developers who contribute more to and participate in discussions in projects with close relationships (i.e., same organization) are more likely to be granted the commit right. \textcolor{black}{This finding is similar with the findings of Bird et al.~\cite{bird2007open}, i.e., social status will positively influence attainment of developer status (becoming committers).} However, different from our qualitative findings, we notice that more offensive comments a developer posted indicate a higher chance of obtaining the commit right ($EXP(Coef) = 1.14$), which is against the communities' code of conduct. It may be because many OSS elites are critical and straightforward, without diluting it with compliments~\cite{Ferreira2021Shut,miller2022did}. For example, a committer in \textit{Node.js} commented --- ``\textit{@*** I'm so angry. I'm so so angry. I hate *** so much sometimes. I hate *** wobbly bullshit with ***}''.

For \textit{Terraform}, the hazard ratios ($EXP(Coef)$) of $\#issue_{org}$ (M10) and $\#issue\_comment_{org}$ (M11) are both greater than one. It indicates that holding the other covariates constant, a higher value of $\#issue_{org}$ and $\#issue\_comment_{org}$ are associated with a higher chance of committer immigration. Thus, issue-related contributions can benefit their chances of entering the trust circle, consistent with the emphasis on committer qualifications of the company-backed model. Similarly, the hazard ratio ($EXP(Coef)$) for $from\_company$ (M9) is 1.70, indicating a strong relationship between the developer's affiliation and increased chances of committer immigration. Holding the other covariates constant, being supported by a company ($from\_company = 1$), increases the hazard by a factor of 1.70, or 70\%. We conclude that developers supported by companies find it easier to obtain the commit right. This is due to its strong commercial involvement --- \textit{Terraform} is created by HashiCorp, also consistent with the findings of RQ1.

As for the negative factors, the p-values for $\#developer$ (control variable) of two projects all approach zero, with hazard ratios $EXP(Coef)$ = 0.23 and 0.19, indicating a strong relationship between the number of developers in the community and decreased chances of becoming committers. It is natural because more people mean more fierce competition and many external developers do not expect to enter the trust circle. For \textit{Node.js}, we also observe that $\#file_{modified}$ (M15) has a slightly negative relationship with committer immigration. This finding is different from our qualitative results. It may indicate that becoming a committer requires developers to deeply understand specific modules, consistent with the modular management of large-scale OSS communities~\cite{tu2000evolution}.

\vspace{1mm}
\begin{mdframed}[linecolor=gray,roundcorner=12pt,backgroundcolor=gray!15,linewidth=3pt,innerleftmargin=2pt, leftmargin=0cm,rightmargin=0cm,topline=false,bottomline=false,rightline = false]
 %, skipabove=10pt, skipbelow=10pt
  \textbf{Summary for RQ2:} We find that the actual selection criteria for committers in practice are basically consistent with community policies. Aspiring developers aiming to gain trust within top OSS communities should possess competitive technical skills from the outset. Initially, they should contribute high-quality code to the project and actively engage in code review activities. Extensive participation in related projects, particularly those within the same organization, can increase the likelihood of gaining community trust. It is important to note the differing emphasis of various OSS governance models on committer qualifications. For instance, in the company-backed model, developers backed by companies may find it comparatively easier to obtain the commit right. However, certain dimensions (e.g., Project Comprehension: \textit{number of files modified}) are not exactly consistent with community policies, and some dimensions (e.g., Assistant: \textit{supporting newcomers}) are not fully evaluated.
\end{mdframed}
\vspace{1mm}

\section{Discussion}\label{discussion}
In this paper, we explore how developers enter the trust circle of top OSS communities. We discuss our findings, the scientific value, the practical implications, and threats to validity below.
\vspace{-0.2cm}
\subsection{Recommendations for Committer Immigration}
The prevailing studies on the immigration of developers, and joining script of developers in particular, mainly focus on how to reduce the contributing barriers of newcomers~\cite{tan2020first,dias2019barriers,steinmacher2015social,steinmacher2015systematic,steinmacher2018almost}. Although some newcomers may successfully onboard through guidance, it is still a great challenge for them to enter the trust circle of OSS communities. We find that only less than 2\% of external developers successfully immigrated. The possibility of obtaining the commit right even gradually decreases since they join the community. This finding is consistent with the prior study, i.e., most newcomers are one-time contributors~\cite{tan2020first,zhou2012make,steinmacher2018almost}. We find that developers' initial performance after they join the communities is critical for being granted the commit right. Initial performance shows concretely as high-quality code contributions, active code review activities, and extensive participation in relevant projects. Thus, for the developers who want to join the community trust circle, the entry barriers may be mitigated if they choose projects that match their skills and actively contribute to standing out quickly.
  
Our findings also suggest that the effect of community cultivation on gaining trust is limited. In modern top OSS communities, external developers lack opportunities to gradually build critical skills for entering the trust circle. It is easy for communities to lose great talent if there are too few opportunities for advancement made available to them. Thus, the trust circle of OSS communities should be friendly to external developers and provide enough time and chance for developer immigration. On the other hand, it also means that the current communities' support for newcomers is insufficient, lacking clear pathways for their growth. Clear pathways allow dedicated developers to understand how they fit into OSS communities long-term, improve engagement and retention, and ultimately, help sustain and accelerate project development. Our findings provide detailed qualifications for gaining trust, which can provide practical guidance for communities to cultivate potential OSS elites. For example, encourage developers to focus on a certain module according to their expertise and interest, actively contribute code, and participate in code review. Future research can focus on the cases of successful committer immigration and investigate how to establish clear and personalized pathways for individual developers.  

\vspace{-0.3cm}
\subsection{Mismatch between Expected and Actual Qualifications}
Although our quantitative analysis basically validates the qualitative findings, the expected qualifications of committers are not exactly consistent with the actual qualifications. This mismatch points out the improvement directions of establishing a scientific mechanism of committer election. For example, many communities ask that nominees should demonstrate exemplary behavior of the code of conduct, e.g., demonstrating empathy and kindness toward other people and being respectful of differing opinions, viewpoints, and experiences. However, through quantitative analysis, we find that more offensive comments a developer posted indicate a higher chance of obtaining the commit right. It implies that these developers are generally straightforward and persistent, and they are probably the authorities in this field. These may be valuable qualities for the head of a massive software project. However, the technical criticisms easily turn into personal attacks and influence the health and safety of online communities~\cite{miller2022did}. To address this issue, it is necessary to strike a balance between technical expertise and positive behavior. Communities should encourage and reward technical excellence while also promoting empathy, kindness, and respect among members. This can be achieved through clear and enforceable codes of conduct, moderation policies, and community guidelines. Additionally, fostering a culture of constructive criticism can help channel technical criticisms in a respectful and productive manner. Encouraging open discussions, providing guidelines for giving feedback, and promoting collaboration can contribute to a healthier community dynamic.

Certain skills may not be adequately evaluated when granting the commit right. For example, many OSS communities expect potential committers to help newcomers actively. However, we do not observe this phenomenon through quantitative analysis. It may be due to the relatively low priority of this skill required by committers or rare opportunities to offer help. 

The above cases reflect the inadequacy of the current committer selection mechanism in practice. The results of RQ1 provide a comprehensive understanding of committer qualifications, which can bring insights for implementing and optimizing committer immigration mechanisms in OSS communities. At the same time, external developers can refer to our findings and try to match the expected qualifications of committers and thus smoothly enter the trust circle of OSS communities.
\vspace{-0.3cm}
\subsection{Threats to Validity}
% defination of committers
In this study, we focus on the developers with the commit right. These developers are usually called ``committers'' in OSS communities, but in some communities, they may refer to ``collaborators'', ``project members'', and ``maintainers''. We acknowledge that these roles have slight differences, e.g., ``maintainers'' are usually used for a project with multiple subsystems.
Because their core commonality is having the commit right, we believe that it is reasonable to analyze the policies of these roles when answering RQ1. 

%limitation of find Committer related files
To determine whether communities have instructions on the commit right, we have read files that may contain this information, including the README file, the documents pointed to by hyperlinks in the README file, the CONTRIBUTING file, and the GOVERNANCE file. However, we acknowledge that there may be a possibility of missing this information during manual searches.

%limitation of manual analysis
For the manual analysis in RQ1, we acknowledge that the choice of some codes is, to some extent, subjective. To mitigate this risk, the extended coding process is performed independently by the second and third authors. Then, they compare the list of codes. The inter-rater reliability during this process was 0.79 (Cohen's Kappa), which indicates substantial agreement between the inspectors and demonstrates the reliability of our coding schema and procedure. Then, based on a coding guide and performing a peer review on each result, the data finally received the full agreement.

%limitation of metrics
To prioritize the different dimensions of committers' qualifications, we developed a set of metrics. However, due to various challenges, such as data accessibility, we were unable to quantify all dimensions, including the ``nomination'' theme. Additionally, some metrics may not fully capture developers' behavior, as our analysis focused solely on their communication behavior on GitHub, while other communication channels like Slack may also be used in communities. Nevertheless, focusing on GitHub is meaningful due to its representativeness and capturing developers' behavior comprehensively presents great challenges, particularly in associating their identities across platforms and identifying specific communications. Our models still reveal multiple significant factors, as evidenced by the p-values obtained from the overall tests. These findings provide valuable insights into committers' qualifications in OSS communities. 
\textcolor{black}{To assess the offensiveness of comments, we employed the \textit{profanity-check} package. Considering that natural language processing tools designed for general domains may not perform well on software engineering datasets, we conducted a random sampling of 384 comments (Confidence Level: 95\%, Margin of Error: 5\%). The results revealed that 362 out of the sampled comments (94\%) were correctly classified. This high accuracy rate indicates the reliability of the tool in identifying offensive language.}
For future studies, it would be beneficial to design alternative metrics and consider behaviors on different platforms, taking into account the taxonomy of committers' qualifications identified in RQ1. This would further enhance our understanding of commit rights in OSS communities.

%limitations of identifying committers
We identify committer immigration through the analysis of the commit log, which cannot guarantee that the identifications are exactly correct in all cases. \textcolor{black}{However, the merge methods employed by the \textit{Node.js} and \textit{Terraform} communities significantly enhance the accuracy of the approach in identifying committer immigration. To evaluate this risk, we can leverage the fact that the \textit{Node.js} community dynamically maintains the README.md file containing the list of current collaborators after November 2017~\cite{Nodejs2024Collaborators}. By comparing our method with the historical records of this list, we can validate the accuracy of our approach. Out of the 62 cases examined, we confirmed 57 cases, indicating a high level of accuracy in our method. In the \textit{Terraform} community, we manually reviewed all 98 immigration cases. Specifically, we analyzed developers’ profiles and community discussions. Out of these, 79 cases were confirmed to have accurate committer information, while 5 cases were incorrectly identified, and 14 cases remained uncertain. These findings demonstrate that our approach for identifying committer information can generally yield reliable results in both the \textit{Node.js} and \textit{Terraform} communities.
It is important to acknowledge that this is currently the most practical method available, and we encourage future studies to explore more precise approaches for identifying committer immigration.} 

%To examine this risk, we randomly sampled 20 committer immigration cases respectively from \textit{Node.js} and \textit{Terraform} and performed manual checks. We analyzed related information,. All 20 cases from \textit{Node.js} were confirmed, and 18 cases from \textit{Terraform} were confirmed. Future studies can try to propose a more precise approach for identifying committer immigration.

% generalization
%We notice that some results are not consistent in these two projects, We acknowledge that some results may not generalize to other OSS communities. However, %as two representative projects in different fields and with different governance models combined with the qualitative findings of RQ1, the credibility of our results is ensured by triangulation verification.
Due to the requirements of survival analysis, we focused on only two projects (\textit{Node.js} and \textit{Terraform}) and did not cover all four types of OSS governance models. However, the community-driven model and the company-backed model, which we did include, accounted for 79\% of the projects. Additionally, we cannot guarantee that these two projects fully represent the two models they belong to. Future studies can adopt our analysis framework to investigate more OSS projects (e.g., different governance models and fields) to expand and complement our findings.
\vspace{-0.1cm}
\section{Conclusion}\label{conclusion}
Committers play a critical role in OSS communities, as they not only demonstrate their exceptional skills but also earn the trust of the community. To gain a comprehensive understanding of how developers establish trust in today's top OSS communities, we conduct an investigation into the community policies regarding the allocation of commit rights. We develop a taxonomy consisting of nine themes to characterize committer qualifications and examine the differences in qualifications across four typical OSS governance models. To further explore the impact of these themes and the practical process of committer selection, we design a set of metrics based on the committer qualifications taxonomy. Through survival analysis, we analyze the actual criteria used for selecting committers. Our findings reveal that the likelihood of committer immigration decreases over time. Developers who consistently produce high-quality code, actively engage in code review, and make extensive contributions to related projects are more likely to become committers. However, we identify a discrepancy between the expected and actual qualifications, suggesting the need for optimization in the distribution process of commit rights. Our study is the first to comprehensively uncover the mechanism of committer immigration in modern top OSS communities, contributing to the growth of external developers and the sustainable development of OSS communities.

\section{Data Availability}\label{sec: data}
We provide the data and scripts online to facilitate replications or future work: \url{https://figshare.com/s/5787e172ff5669b2ccb3}.

\section{Acknowledgements}
This work is supported by the National Natural Science Foundation of China Grants 62202022, 62141209, and 62332001.
%\balance
\bibliographystyle{ACM-Reference-Format}
\bibliography{sample-base}

\end{document}